\documentclass[showpacs,pre,aps,twocolumn]{revtex4-1}

\pdfoutput=1
\usepackage{amsmath,amsfonts,amssymb,bm,graphicx,hyperref,color,units}

\newcommand{\OO}{\mathcal{O}}
\newcommand{\Com}{\mathbb{C}}
\newcommand{\ZZ}{\mathbb{Z}}

\newcommand{\bra}[1]{\langle #1|}
\newcommand{\ket}[1]{|#1\rangle}
\newcommand{\Tr} {{\text T}{\text r}}
\newcommand{\vlam}{\vec{\lambda}}

\begin{document}
\title{Dynamical phase transitions, time-integrated observables and
geometry of states}
\author{James M. Hickey}
\author{Sam Genway}
\author{Juan P. Garrahan}
\affiliation{School of Physics and Astronomy, University of Nottingham,
Nottingham, NG7 2RD, United Kingdom}

\date{\today}

\begin{abstract}

We show that there exist dynamical phase transitions (DPTs), as defined in [Phys. Rev. Lett. \textbf{110} 135704 (2013)], in the transverse-field Ising model (TFIM) away from the static quantum critical points.  We study a class of special states associated with singularities in the generating functions of time-integrated observables found in [Phys. Rev. B \textbf{88} 184303 (2013)].  Studying the dynamics of these special states under the evolution of the TFIM Hamiltonian, we find temporal non-analtyicities in the initial-state return probability associated with dynamical phase transitions.
By calculating the Berry phase and Chern number we show the set of special states have interesting geometric features similar to those associated with static quantum critical points.
\end{abstract}

\maketitle

\section{Introduction}
\label{sec:Intro}

Phase transitions are remarkable phenomenon which are ubiquitous in nature and have been studied in equilibrium thermodynamics since the 19$^\text{th}$ century~\cite{Maxwell1875}.  Recently, driven by experimental advances~\cite{Polkovnikov2011,Bloch2002,*Weiss2006,*Gring2012,*Cheneau2012}, much interest has turned to the study of non-equilbrium dynamics and phase transitions ~\cite{Polkovnikov2011,Heyl2013,Karrasch2013,Pollmann2010,Gambassi2012,Calabrese2011,*Calabrese2012,*Calabrese2012b,*Essler2012}. 
Recent work by  Heyl, Polkovnikov and Kehrein~\cite{Heyl2013}
revealed an interesting connection between non-analyticities in the
non-equilibrium dynamics of a quantum system and the theory of equilibrium phase
transitions. The authors highlighted the formal similarities between the appearance of non-analyticities in the return amplitude of a quantum system and the Lee-Yang theory of equilibrium phase transitions~\cite{Lee1952,*Yang1952,Fisher1965}.  Studying in detail the transverse-field Ising model (TFIM), the authors studied quantum quenches across boundaries between the quantum phases as well as quenches within a quantum phase.  Only when quenching across a phase boundary were nonanalyticities revealed in the temporal behaviour of the return amplitude.  In the thermodynamic limit, temporal nonanalyticities emerged due to the coalescence of Lee-Yang zeroes in the complex plane of the return amplitude.  Due to strong similarities with equilibrium phase transitions they have been dubbed dynamical phase transitions (DPTs)~\cite{Heyl2013}.

\begin{figure}
\includegraphics[width=1.0\columnwidth]{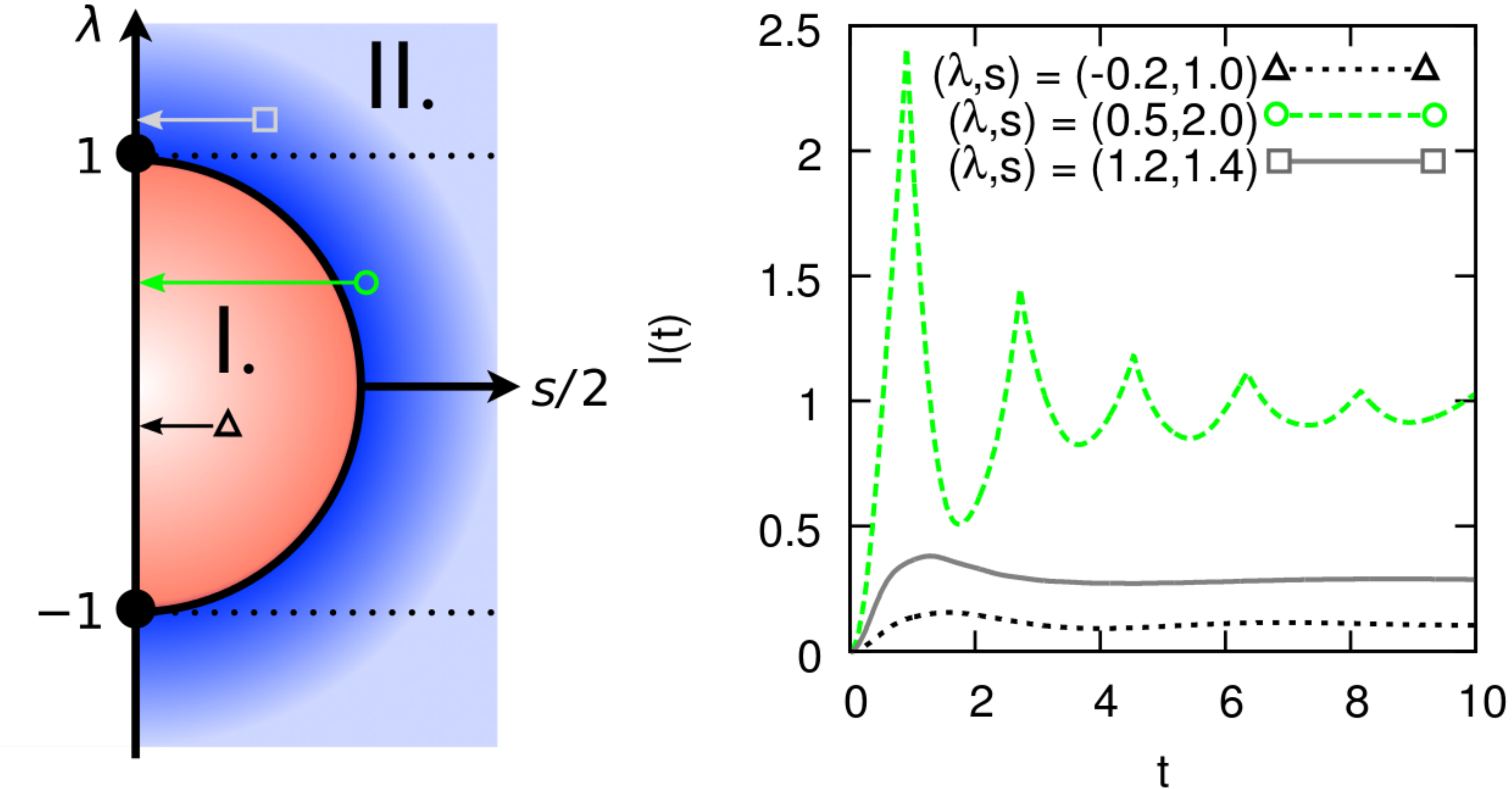}
\caption{ In the left panel is the FCS phase diagram of the TFIM, regions I and II are the
dynamically ordered and disordered regimes respectively~\cite{Hickey2013}.  The counting field is labelled $s$ and $\lambda$ is the transverse magnetic field strength. We
consider ``quenches'' from points $(\lambda,s) \rightarrow (\lambda,0)$.  The right
panel shows the large deviation function associated with the return probability
of this protocol.  Dynamical non-analyticities are found when the ``quench''
crosses the FCS critical line, analoguous to the effects of a quantum quench
across a static quantum critical point. \label{fig:fig1}}
\end{figure}

Other studies, by us and others, have found dynamical insights into many-body dynamics by exploring time-integrated observables~\cite{Ruelle2004,Lecomte2007,Garrahan2007}.  Making use of full counting statistics (FCS) methods~\cite{Levitov1993,*Levitov1996, Nazarov2003,*Nazarov2003b, Pilgram2003, Flindt2008, Garrahan2007,Esposito2009, Lecomte2007,Flindt2009},
the moment generating function (MGF) for time-integrated observables is treated analogously to a partition
sum.  In this so-called `$s$-ensemble' approach, the status of the counting field `$s$' is elevated to that of a thermodynamic variable~\cite{Hedges2009, Pitard2011, *Speck2012}.  Pursuing the thermodynamic analogy, singular features
in the long-time behaviour of the cumulant generating function (CGF) have been identified as phase transitions in the FCS~\cite{Ates2012,Garrahan2010,Genway2012, Hickey2012}.  In a recent paper~\cite{Hickey2013}, we studied moments of the time-integrated transverse magnetisation in the TFIM with this formalism and uncovered a set of FCS singularities [see Fig.~\ref{fig:fig1}(left)] in the CGF.  Analogous to the ground state at the static quantum critical points in the model, there exist a special class of states which capture the singular FCS features.  These are eigenstates of a the non-Hermitian operator which forms the MGF.

In this paper, we explore this special class of states in greater detail.  We study their evolution under the TFIM Hamiltonian and find DPTs emerge, similar in nature to those uncovered by Heyl, Polkovnikov and Kehrein in Ref.~\cite{Heyl2013}.  Therefore, we demontrate DPTs can exist \emph{without} performing a quantum quench across a static quantum critical point in the TFIM.  In such cases, we find, as in Refs.~\cite{Heyl2013,Karrasch2013,Fagotti2013}, nonanalyticities in the initial-state return probability [see Fig.~\ref{fig:fig1}(right)].  We develop further the relation of the special states with groundstates near quantum criticality by exploring the geometric properties~\cite{Berry1984,Simons1983, Aharonov1987, Samuel1988, Wu1988} of the class of special states.   We find the geometric properties of these states at the FCS singularities exhibit features similar to the ground state at a static quantum critical point~\cite{Ma2009,Polkovnikov2013b,Hawking1983}.

The paper is organized as follows: in Sec.~\ref{sec:FCS} we provide the
theoretical framework on time-integrated observables and FCS criticality in
closed systems.  We then outline its application to the TFIM in Sec.~\ref{sec:TFIM} and its connection to the return probability in Section~\ref{sec:squench}.
In Sec.~\ref{sec:GP} we provide an overview on the geometric-phase
characterization of states and its application to DPTs.  We then present our results concerning the connection between these FCS phases and DPTs in Sec.~\ref{sec:Res1} before discussing the
geometry of the states associated with these phases in Sect.~\ref{sec:Res2}. 
Finally, we present our conclusions in Sec.~\ref{sec:Conc}.

\section{Theoretical Formalism}
\subsection{Generating Functions and Time-Integrated Observables}
\label{sec:FCS}

Central to the theory of phase transitions is the partition function for a system
\begin{equation}
Z(\beta) = \Tr(e^{-\beta H}) = e^{-N\beta f(\beta)}
\label{eq:Z}
\end{equation}
where $\beta$ is the inverse temperature.  Here, $f(\beta)$ is the free energy density, in which nonanaltics associated with phase transntions manifest, and $N$ is the number of degrees of freedom.  As noted in~\cite{Heyl2013}, if the system is in a pure state, the partition function is related to the quantity
\begin{equation}
 G(t) = \bra{\psi}{e}^{-iHt}\ket{\psi}\,.
 \label{eq:G}
\end{equation}

In the context of a quantum quench~\cite{Gambassi2012,Polkovnikov2011} $G(t)$ is the Loschmidt amplitude~\cite{Silva2008}, with $H$ the post-quench Hamiltonian and $\ket{\psi}$ the initial state.  Mapping $it \rightarrow \beta$, for $\beta \in \Com$, the Loschmidt amplitude is now the boundary partition function
$Z(\beta) = \langle {e}^{-\beta H}\rangle$.  This partition function has zeros in the complex $\beta$-plane which lie on the real time axis when ${e}^{-i H t}\ket{\psi}$ is orthogonal to $\ket{\psi}$. In the thermodynamic limit these zeros coalesce and may appear as nonanalyticities in the rate function, $l(t)$, for the return probability
\begin{equation}
l(t) =\lim_{N\rightarrow \infty} \frac{1}{N}\log |G(t){|}^{2}
\label{eq:l}
\end{equation}

The partition function, and Loschmidt amplitude, are generating functions for
the energy of the system and work done during a quench.  These are both static quantities.  Singular features in the free-energies of these generating
functions correspond to quantum phase transitions and dynamical phase transitions respectively. 
We now look at a generating function for purely \emph{dynamical} quantities, namely time-integrated observables~\cite{Hickey2013}.  Consider a closed quantum system with a Hamiltonian, $H$.  We wish to examine moments of a time-integrated observable
\begin{equation}
K_{t} \equiv \int^{t} k(t') \text{d}\text{t}'
\end{equation}
where $k(t')$ is the operator associated with the observable of interest written in the Heisenberg
representation.  The MGF of this quantity is directly related to a 
non-Hermitian Hamiltonian, $H_{s}$, and an associated non-unitary evolution
operator $T_{t}(s)$, defined by
\begin{equation}
T_{t}(s) \equiv {e}^{-i t H_{s}}, H_{s}\equiv
H-\frac{\text{i}\text{s}}{2} k\,.
\end{equation}
With these operators one can show that the MGF of $K_{t}$ is given by
\begin{equation}
\label{eq:Gen}
Z_{t}(s) = \langle T^{\dag}_{t}(s) T_{t}(s) \rangle
\end{equation}
 and moments of $K_{t}$ are generated through its derivatives, $\langle
 K_{t}^{n} \rangle = (-)^{n} \partial_{s}^{n} Z_{t}(s) |_{s \to 0}$, while the logarithm of the MGF, 
 $\Theta_{t}(s) \equiv \log Z_{t}(s)$, is the CGF.   This forms the definition of a form of FCS~\cite{Nazarov2003,Nazarov2003b}, where in contrast to the usual definition of FCS we take the parameter $s$ to be real. To study the analytic properties of this generating function it will be useful to study the associated scaled CGF in the long-time limit
 \begin{equation}
 \label{eq:scaling}
 \theta(s) = \lim_{N, t\rightarrow \infty}
 \frac{\Theta_{t}(s)}{Nt}.
 \end{equation}
 
 \subsection{Application to the Transverse-Field Ising model}
 \label{sec:TFIM}
 We focus on the one-dimensional TFIM with 
 periodic boundary conditions, whose Hamiltonian is given by
 \begin{align}
 \label{eq:Ham}
 H &= -\sum_{i} \sigma^{z}_{i}\sigma^{z}_{i+1} -\lambda \sum_{i}\sigma^{x}_{i}
 \nonumber \\
 &= \sum_{k} \epsilon_{k}(\lambda)(\gamma^{\dagger}_{k}\gamma_{k}-1/2)\,,
 \end{align}
where $\sigma^{x,z}$ are pauli matrices and $\lambda$ is the strength of the magnetic field.  The Hamiltonian is diagonalized using a Jordan-Wigner transformation
 followed by a Bogoliubov transformation~\cite{Sachdev2011} and the spectrum   (see Appendix) is
 \begin{equation}
 \epsilon_{k}(\lambda) = 2\sqrt{(\lambda-\cos k)^2 +\sin^2 k}\,.
 \label{eq:ek}
 \end{equation}
We examine the generating function associated with the
 time-integrated transverse magnetization, $\sigma^{x} = \sum_{i}
 \sigma^{x}_{i}$ when the system is in the ground state.  Using standard free-fermion techniques (see Appendix) one finds the CGF~\cite{Hickey2013}
 \begin{equation}
 \theta(s) = 4\text{I}\text{m}\Bigl(\int^{\pi}_{0} \Bigl|\sqrt{(\lambda+is/2 -
 \cos k)^{2} +\sin^{2}k}\Bigr|~dk\Bigr).
 \end{equation}
 Making an analogy to equilibrium statistical mechanics we treat this CGF as a type of dynamical ``free energy" and introduce an order parameter,
$-\theta'(s)$, and a \emph{dynamical} susceptibility, $\chi_{s} = \theta''(s)$, where $'$
denotes differentiation with respect to $s$. Using these dynamical quantities the FCS
phases can be characterised [see Fig.~\ref{fig:fig1}].  A whole critical line exists where $\chi_{s}$ diverges~\cite{Hickey2013}.  The static quantum critical points lie at the end of this critical line.  This critical curve, shown in Fig.~\ref{fig:fig1}, is a circle in the $\lambda-s$-plane defined by
 \begin{equation}
 \label{eq:curve}
 \lambda^{2} + (s/2)^2 = 1.
 \end{equation}
 
   The critical line~\eqref{eq:curve} corresponds to a closing in the gap of the complex spectrum
of $H_{s}$ at a particular wavevector, $k_{\lambda}$, which depends on the transverse magnetic field.  For $|\lambda| < 1$ the critical $s$ value is given
by 
\begin{equation}
s_{c} = 2\sin k_{\lambda} \,;\quad  k_{\lambda} = \cos^{-1}\lambda \,.
\label{eq:k_lam}
\end{equation}
 This phase diagram is divided into two regions, region I and II (see
Fig.~\ref{fig:fig1}), which we will refer to as `dynamically ordered' and
`dynamically disordered' respectively.  Associated with each point in this FCS
phase diagram we may associate a state $\ket{s}$ defined by
$|s\rangle \equiv \lim_{t \to \infty} T_{t}(s) | i \rangle$~\cite{Hickey2013}, for initial states $\ket{i}$~\footnote{This state is independent of the initial state provided the initial state has finite overlap with it.} with an appropriate normalisation.  The states $\ket{s}$ are right eigenstates of $H_s$.  In our case the initial state is the vacuum of the TFIM.  With this choice $\ket{s}$ takes different forms depending on the values of $\lambda$ and $s$:
\begin{align}
\label{eq:sprod1}
\ket{s} &= \bigotimes_{k>0} \ket{s_{k}} \\ 
&\propto \left\{
\begin{array}{lc}
\bigotimes_{k>0} {\ket{1_k,1_{-k}}}_{s} & \lambda>1 \\ \nonumber
\\
\bigotimes_{k<k_{\lambda}} {\ket{0_k,0_{-k}}}_{s} 
\bigotimes_{k>k_{\lambda}} {\ket{1_k,1_{-k}}}_{s} & -1<\lambda<1 \\ \nonumber
\\
\bigotimes_{k>0} {\ket{0_k,0_{-k}}}_{s} & \lambda<-1 \\ \nonumber
\end{array}
\right. 
\end{align}

The states $\ket{n_{k},n_{-k}}_{s}$ are eigenstates of $H$ with $n_k$ ($n_{-k}$) fermions in the mode $k$ ($-k$) (see Appendix).  It turns out that the states $\ket{s}$ can be prepared to high precision by coupling the system to a simple Markovian environment~\cite{Hickey2013}.

\subsection{Quenches in $s$}
\label{sec:squench}
In this paper, we will consider the following quench protocol:  we initially connect the Ising
chain to an appropriate bath (as in~\cite{Hickey2013}) and allow it to evolve towards the state $\ket{s}$.  After this we will
perform a ``quench'' in the $s$ parameter by decoupling the system from the environment and evolving the state $\ket{s}$ under the TFIM Hamiltonian $H$.  

\begin{equation}
\label{eq:sprod2}
\ket{s_{t}} = {e}^{-i H t}\ket{s} = \bigotimes_{k>0} {e}^{-i H t}\ket{s_{k}} = \bigotimes_{k>0} \ket{s_{k,t}}
\end{equation}

In this scheme, which we will refer to as the `$s$-quench', we will examine
how dynamical phase transitions are related to the critical FCS line shown in Fig.~\ref{fig:fig1}.  Furthermore we will
characterize the geometric properties of the states $\ket{s}$ ($\ket{s_{t}}$), expressed in terms of the
majorana fermions of $H$, focussing on the Berry phase (Chern number).  This approach has previously been used to
characterize ground state properties, quantum phase transitions and DPTs:  here, we will
use it to characterize the new FCS critical line using states $\ket{s}$.  We provide a brief discussion of these geometric parameters so we may link these to both quantum as well as FCS phase transitions and connect these critical features with DPTs.

\begin{figure*}[th]
\includegraphics[width=1.2\columnwidth]{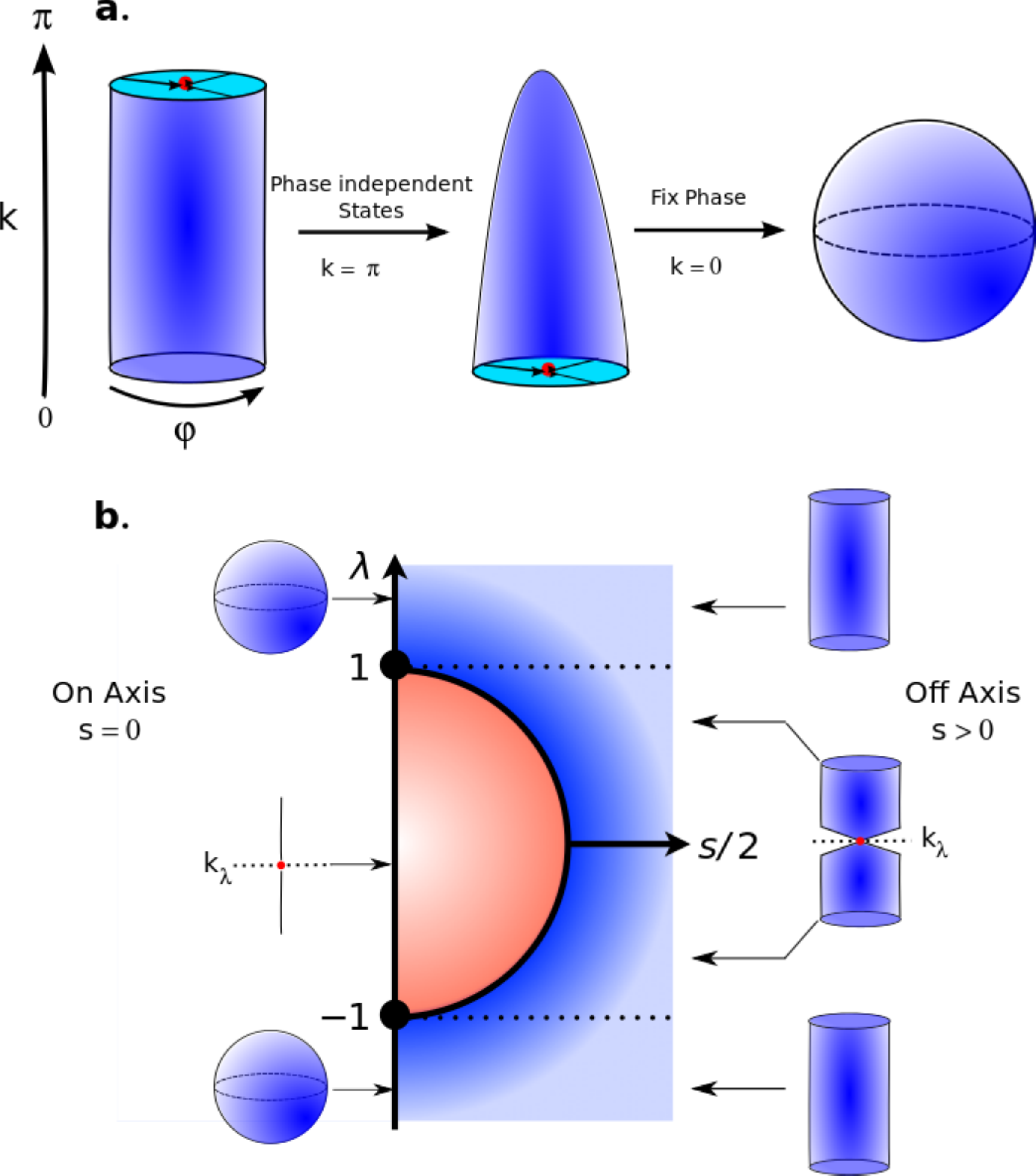}
\caption{ (a) The parameter manifold $M^2$ is topologically equivalent to the
$S^2$-sphere as the $k=\pi$ mode is $\varphi$ independent and the $k = 0$ mode is
(at most) only dependent on $\varphi$ up to a gauge transformation. (b) Examining
the manifold at each point in the  FCS phase diagram we find that at $k_{\lambda}$
the manifold is $\varphi$ independent. The $s$-states below $|\lambda| = 1$ are
found to be completely $\varphi$ independent on the $s = 0$ line.
However above the $|\lambda| = 1$ line the manifold once again becomes
topologically equivalent to the $S^2$-sphere at $s = 0$.  
\label{fig:fig2}}
\end{figure*}

\subsection{Geometric Phase and Berry Curvature}
\label{sec:GP}

Topological quantum numbers provide an alternative way of
classifying and characterizing the ground state properties of many-body quantum
systems~\cite{Polkovnikov2013b}.  This geometric approach to studying ground state properties of
many-body systems has provided a new interpretation of quantum phase
transitions~\cite{Carollo2005,Zhu2006,Hamma2006}.  One of the most widely-used measures of these geometric properties
of physical systems is the Berry phase~\cite{Berry1984} associated with adiabatic transport
of quantum state vectors around a closed parameter-manifold.  Associated with this phase
we may construct an associated Berry curvature, which when we consider transport
along a two-dimensional manifold, $M^{2}$, leads to the Chern number of the
system.  We will now discuss these quantities in more detail, before 
describing their relationship to results on dynamical quantum phase
transitions.

Consider a manifold of Hamiltonians defined by some parameters
$\vlam$.  A natural measure of the distance~\cite{Provost1980} between the
ground states, $\ket{0(\vlam)}$, of this manifold is 
\begin{equation}
ds^{2} = 1-|\bra{0(\vlam)}0(\vlam + d\vlam)\rangle| =
\sum_{\mu,\nu}g_{\mu\nu}d\lambda^{\mu}d\lambda^{\nu}
\end{equation}
where $g_{\mu\nu}$ is the geometric tensor
\begin{equation}
g_{\mu\nu} = \bra{0(\vlam)}\overleftarrow{\partial_{\mu}}
\partial_{\nu}\ket{0(\vlam)} -
\bra{0(\vlam)}\overleftarrow{\partial_{\mu}}\ket{0(\vlam)}\bra{0(\vlam)}\partial_{\nu}\ket{0(\vlam)}.
\end{equation}
Here $\partial_{\mu} = \partial/\partial \lambda^{\mu}$ and the diacritic arrow, $\overleftarrow{}$, indicates the partial derivative acts to the left.  Now the imaginary part of the geometric tensor is directly related to the Berry
curvature, $F_{\mu\nu}$, via :
\begin{equation}
F_{\mu\nu} = -2\text{I}\text{m}[g_{\mu\nu}] =
\partial_{\mu}A_{\nu}-\partial_{\nu}A_{\mu}
\end{equation}
with $A_{\mu} = i\bra{0(\vlam)}\partial_{\mu}\ket{0(\vlam)}$ the Berry
connection.  The line integral of this connection is simply the Berry phase
($B$) and the surface integral, over the parameter manifold ($\mathcal{M}$), of the curvature is related to a quantity known as the
Chern number, $C$:

\begin{align}
\label{eq:Berry}
&B \equiv \int_{\partial S}\vec{A}.d\vlam \nonumber \\
& C \equiv \frac{1}{2 \pi} \int_{\mathcal{M}} F_{\mu\nu} dS_{\mu\nu} .
\end{align}

  If the manifold is closed in the topological sense, then the Chern number is simply an
integer~\cite{Thouless1994,*Frankel2004}.  We now summarise existing results on two-parameter manifolds which are of relevance to the TFIM~\cite{Heyl2012a}.  Consider the evolution of the ground state of the
TFIM at $\lambda = \lambda_{i}$  after a quench to $\lambda_{f}$.  The time evolved state factorizes into contributions from each momentum sector, $\ket{u_{k,t}}$,
\begin{equation}
\label{eq:ukt}
\ket{u_{k,t}} = [\cos \phi_{k} - i \sin \phi_{k}
{e}^{-2i\epsilon_{k}(\lambda_{f})t}\gamma_{k}^{\dagger}\gamma_{-k}^{\dagger}] \ket{0_{k},0_{-k}}.
\end{equation}
The annihilation operators $\gamma_k$ diagonalise the TFIM Hamiltonian [see Eq.~\eqref{eq:Ham}], with $\epsilon_k$~\eqref{eq:ek} the excitation spectrum. The angles $\phi_{k}$ are the Bogoliubov angles associated with the Bolgoliubov transformation employed in diagonalising $H$, and
$\ket{0_{k},0_{-k}}$ is the vacuum associated with the mode $k$ ($\gamma_{\pm k} \ket{0_{k},0_{-k}} = 0$). 
Examining Eq.~\eqref{eq:Ham}, one can see that the final Hamiltonian obeys a
global $U(1)$ symmetry

\begin{equation}
\label{eq:pshift}
\gamma_{k} \rightarrow \gamma_{k}{e}^{-i\varphi},
\end{equation}
however this symmetry is not shared by the time evolved state due to the
spontaneous creation of excitations~\cite{Heyl2012a}.  Furthermore this geometric phase corresponds to the global phase accumulated on cyclic
evolution from $\varphi = 0$ to $\pi$.  

We now focus on the manifold of states, $M^{2}$, defined by parameters $k$ and $\varphi$.  The parameters $k$ and
$\varphi$ are defined on the same interval $[0,\pi]$; this can be seen from
the from the form of $\ket{u_{k,t}(\varphi)}$.  These states are defined
uniquely for $k > 0$ and $k < \pi$.  Furthermore as the excitations are Cooper-pair-like in nature (see Eq.~\eqref{eq:ukt}) the phase factor $\varphi$ has a factor of $2$ preceding it.  This implies the states are uniquely defined
with $\varphi \in [0,\pi]$.  Now in order to compute the curvature of $M^2 =
 [0,\pi]\times[0,\pi]$, we have to compute the derivative of $\ket{u_{k,t}(\varphi)}$ with respect to both $\varphi$ and $k$.  The derivative with respect to $\varphi$ is straightforward, while the derivative with respect to $k$ is less so.  In order to compute this we need to use perturbation theory and expand $\ket{0_{k},0_{-k}}$ as follows:
\begin{equation}
\ket{0_{k+\delta k}, 0_{-(k+\delta k)}} = \ket{0_{k},0_{-k}} +
\frac{d}{dk}\ket{0_{k},0_{-k}} \delta k + \OO(\delta k^{2}).
\end{equation}
Standard perturbation theory has an implicit gauge choice built in known as
 the parallel transport gauge:  because the 
 excited states are orthogonal to the ground state, it turns out~\cite{Resta2000} 
 \begin{equation}
 \frac{d}{dk}\ket{0_{k},0_{-k}} =
 \frac{d}{dk}\gamma^{\dagger}_{k}\ket{0_{k},0_{-k}} = 0,
 \end{equation} 
 making the calculation of the Berry Phase and curvature straightforward.
 This manifold was examined with these states in~\cite{Heyl2012a} where it was
 found that the states $\ket{u_{k,t}(\varphi)}$, at $k = \pi$, are independent of the phase
 $\varphi$, i.e. $\sin \phi_{\pi} = 0$.  This behaviour is independent of the quench protocol.  However, the behaviour of the
 infrared states ($k\rightarrow 0$) is slightly more complex. If we quench within the same phase
 the $k = 0$ states are $\varphi$ independent, as $\sin \phi_{k\rightarrow 0} = 0$, and so the manifold $M^{2}$ is topologically equivalent to the 2-sphere.
 However, quenching across the static quantum critical point we find the $k = 0$
 states now depend on $\varphi$ up to a global phase, as $\sin
 \phi_{k\rightarrow 0} = 1$.
 This global phase may be removed with an appropriate gauge transformation, once again leading to
 topological equivalence with a 2-sphere (see Fig.~\ref{fig:fig2}).  This
 dependence on $\varphi$ up to a total phase is due to a population inversion, of the new modes associated with the quench Hamiltonian, on
 crossing the critical point and is captured by a change in the Chern number,
 which is given by $C = \sin^2 \phi_{k=0}$.
 For quenches within the same phase $C=0$ but for quenches across the
 critical point it was found $C=1$, this change in topological number
 corresponds to the emergence of dynamical phase
 transitions~\cite{Heyl2012a}.

In this work we apply this idea of geometric phase using the set of states
$\ket{s_{k,t}}$, defined in Eqs.~\eqref{eq:sprod1} and \eqref{eq:sprod2}, in place of the $\ket{u_{k,t}}$ states.  By studying the
dynamical properties of our $s$-quench and the geometric properties of the $s$-states we will extend the links between the geometric phase and both static and dynamical criticality to the FCS
critical line in the TFIM.

\section{Results}

\subsection{The $s$-Quench}
\label{sec:Res1}
Having introduced the necessary background to DPTs and the extended set of $\ket{s}$
states we are interested in, we now examine the features of $\ket{s}$ states
under the evolution of the TFIM Hamiltonian.  From
Eqs.~\eqref{eq:Ham} and~\eqref{eq:sprod1} we
find that the rate function associated with the return probability (Eq.~\eqref{eq:l}) takes the form
\begin{align}
\label{eq:LE}
l(t) = 2\text{R}\text{e}&\Bigl(\int^{k_{\lambda}}_{0} \log \Bigl[
\frac{|\cos \alpha^{s}_{k}|^{2} +|\sin
\alpha^{s}_{k}|^{2}e^{-2i\epsilon_{k}t}}{\cosh
2\text{I}\text{m}(\alpha^{s}_{k})} \Bigr]dk \Bigr.
\nonumber \\
&\Bigl.+\int^{\pi}_{k_{\lambda}} \log \Bigl[
\frac{|\sin \alpha^{s}_{k}|^{2} +|\cos
\alpha^{s}_{k}|^{2}e^{-2i\epsilon_{k}t}}{\cosh
2\text{I}\text{m}(\alpha^{s}_{k})} \Bigr]dk  \Bigr)
\end{align}
with $k_\lambda$ defined in Eq.~\eqref{eq:k_lam}.  In the finite-$N$ regime this function will contain zeros at times
 \begin{equation}
 t_{n} = \frac{i}{(2)\epsilon_{k}}(\log |\tan \alpha^{s}_{k}|^{2} + i(2j +
 1)\pi)\,.
 \end{equation}
Here $j \in \ZZ$ and the $(2)$ indicates there are two sets of zeros.  One set
is due to the integrand in Eq.~\eqref{eq:LE} vanishing; the other is
attributed to the integrands attaining the same value and emergent
non-analytic behaviour at the $k_{\lambda}$ limits of these integrals. In
both cases these zeros lie on the real time axis whenever the complex angle, $\alpha^{s}_{k}$ (see Appendix), is such that $|\cos \alpha^{s}_{k}| = |\sin \alpha^{s}_{k}|$.
This is only ever the case when the initial $\ket{s}$ lies in the dynamically
disordered phase and $|\lambda| < 1$: in this region there is a well defined $k_{\lambda}$ and the `quench' crosses the FCS critical line described by Eq.~\eqref{eq:curve}.  Within
this regime, in the thermodynamic limit, we see the emergence of non-analyticities at critical times  
\begin{equation}
t_{*} = \frac{(2j+1)\pi}{2\epsilon_{k_{\lambda}}},
\frac{(2j+1)\pi}{\epsilon_{k_{\lambda}}}\,.
\end{equation}

  When we prepare the system in a state $\ket{s}$ from within the
dynamically ordered regime such that we don't
`quench' across the critical line (see Fig.~\ref{fig:fig1}), or at $|\lambda| > 1$ where there is no critial line,
no DPTs are visible; in this region there is no energy scale $k_{\lambda}$.
The interpretation of this new energy scale set by $k_{\lambda}$ is simple:  the occupation of
this mode $n_{s=0}(k_{\lambda}) = 1/2$.  The other modes have occupation $n_{s=0}(k) < 1/2$ for
$k<k_{\lambda}$ and $n(k) = 1/2$ for $k > k_{\lambda}$, so we see $k_{\lambda}$
marks the onset of half-occupancy. These results are 
similar to those in Ref.~\cite{Heyl2013} for ground states
with one crucial difference. The emergence of these DPTs is not due to quenching across a static quantum critical
point~\cite{Heyl2013,Karrasch2013,Pollmann2010}, the existence of the
infinite temperature critical mode was built into the states $\ket{s}$, see
Eq.~\eqref{eq:sprod1}.  This mode defines the critical features in the FCS of
the time-integrated magnetization of this system and so by creating states which capture such singularities we build in a critical mode $k_{\lambda}$.  
 
To summarise our results in this subsection, we have shown for the a particular choice of initial state which is not a ground state of the TFIM, 
DPT features emerge even far from quantum criticality.   We propose that one could even
consider the FCS critical line as a critical line for these extended set of states. To corroborate this idea of a new line of criticality we will now
examine the geometric properties of these states.

\subsection{Geometry of $s$-states}
\label{sec:Res2}

Now we turn to a geometric characterisation of the states $\ket{s}$~\eqref{eq:sprod1} and $\ket{s_{t}}$~\eqref{eq:sprod2}, beginning with the an analysis of the former.  These states, $\ket{s}$, are the right eigenvectors of $H_{s}$.  Previously it has been shown~\cite{Polkovnikov2013b,Carollo2005,Zhu2006,Hamma2006} that the geometry of a system's ground state shows signatures of static quantum critical points.  By analogy we expect that the geometric properties of these new states, $\ket{s}$, should show signatures of the FCS critical line~\eqref{eq:curve}.  However, as this is not a conventional quantum critical line it is unclear exactly how these geometric quantities such as the Berry Phase and Chern number will behave in its vicinity.  We begin by introducing a family of TFIM Hamiltonians, $H(\varphi)$, which depend on the parameter $\varphi$ associated with the global phase shift of fermionic operators~\eqref{eq:pshift}.
Diagonalizing both $H_{s}$ and $H$ we are able to express the FCS-critical state $\ket{s}$ in terms of the fermionic modes of the final Hamiltonian $H$. Applying a global phase
shift, Eq.~\eqref{eq:pshift}, to the fermionic operators of $H$, we obtain a new state
$\ket{s(\varphi)}$, see Appendix.
From Eq.~\eqref{eq:Berry}, the Berry phase associated with this adiabatic evolution is given by
\begin{equation}
\label{eq:BerryCalc}
B = \int^{\pi}_{0} \bra{s(\varphi)}i\partial_{\varphi}\ket{s(\varphi)} d\varphi.
\end{equation}
To study the FCS critical line it is necessary to work in the thermodynamic
limit where the size of the spin chain, $N\rightarrow \infty$.  In this limit
we study the geometric density, $\tilde{\beta} \equiv \lim_{N\rightarrow \infty}B/N$,
and inserting $\ket{s(\varphi)}$ in Eq.~\eqref{eq:BerryCalc} we find it takes the form
\begin{equation}
\tilde{\beta} = -\int^{\pi}_{k_{\lambda}} \frac{|\cos \alpha^{s}_{k}|^{2}}{\cosh
2\text{I}\text{m}(\alpha^{s}_{k})} dk - \int^{k_{\lambda}}_{0}
\frac{|\sin \alpha^{s}_{k}|^{2}}{\cosh 2\text{I}\text{m}(\alpha^{s}_{k})} dk.
\end{equation}

\begin{figure}[h!]
\includegraphics[width = 1.0\columnwidth]{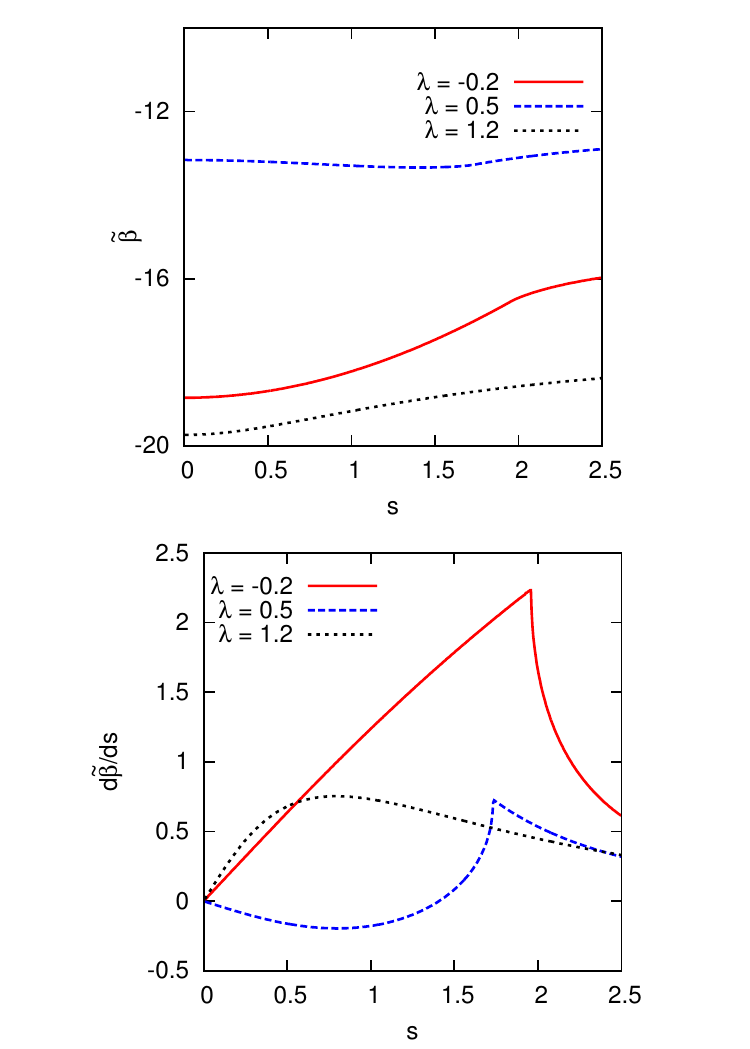}
\caption{The top panel shows slices of the Berry phase density at various
$\lambda$ values.  No extremum or singular features are visible in the vicinity
of the FCS critical line, see Fig.~\ref{fig:fig1}.  However, the derivative of this density,
d$\tilde{\beta}$/d$s$, has extremum located at the FCS critical line, this is shown in
the bottom panel.
For $|\lambda| > 1$ no such extremum are present, this is due to a lack
of any FCS critical points in this parameter regime.
\label{fig:fig3}}
\end{figure}

In Fig.~\ref{fig:fig3} we plot $\beta$ for some representative slices through the ($\lambda,s$)-plane.  The geometric phase
density appears to be the same in both regions I
and II of the FCS phase diagram, see Fig.~\ref{fig:fig1} and Fig.~\ref{fig:fig3}.  However on examining $d\tilde{\beta}/ds$,  minima
appear at the FCS critical line.  Previously such extrema were
used as a method to identify quantum
criticality~\cite{Ma2009,Carollo2005,Zhu2006,Hamma2006} but here they also mark the critical features not of the final Hamiltonian but of $H_{s}$.  

\begin{figure}
\includegraphics[width = 1.0\columnwidth]{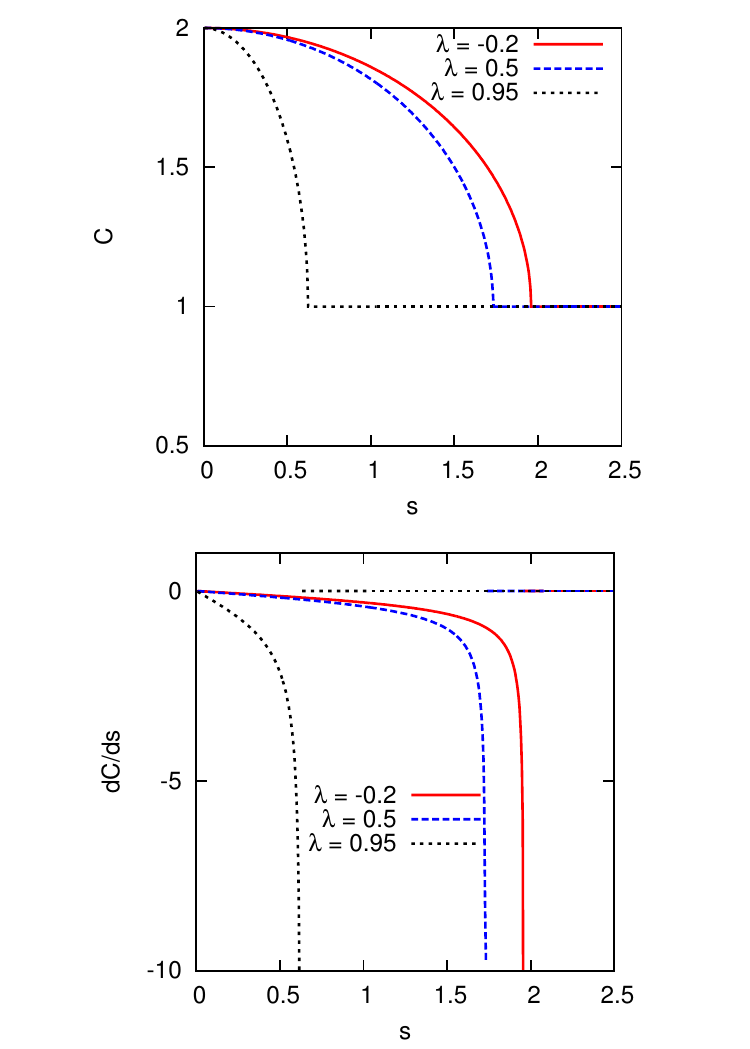}
\caption{The Chern number, $C$, associated with the manifold of $\ket{s_{k,t}}$
states has a ``kink" at the FCS critical line, this is shown for
various $\lambda$ slices in the top panel.  The derivative of the Chern number, displays
divergences at the FCS critical line, these features are usually associated with
quantum criticality but now mark this new FCS critical line, bottom panel.
\label{fig:fig4}}
\end{figure}

We have shown that the Berry phase can be used as a method to identify FCS
critical points.  We note that this is due to the connection of the Berry phase and the energy-level structure of $H_{s}$.  To connect this observation with the emergence of
DPTs we perform an analysis of the Chern number, $C$, discussed in
Sec.~\ref{sec:GP}.  Still working in the thermodynamic limit we now consider the ``quenched" state $\ket{s_{t}(\varphi)}$, including the fermionic shift ~\eqref{eq:pshift}, and split it up into its contributions from each momentum sector $k$.
Combining the result with Eq.~\eqref{eq:Berry}, the Chern number has a highly non-trivial functional form
\begin{align}
&C = \frac{2(|\cos \alpha^{s}_{k_{\lambda}}|^{2} - |\sin
\alpha^{s}_{k_{\lambda}}|^{2})}{\cosh 2\text{I}\text{m}(\alpha^{s}_{k_{\lambda}})} \nonumber \\
&+\text{I}\text{m}\Bigl(\int^{k_{\lambda}}_{0} \frac{2i (\sin
\alpha^{s}_{k})^{*}\cos \alpha^{s}_{k}\partial_{k}\alpha^{s}_{k}}{\cosh
2\text{I}\text{m}(\alpha^{s}_{k})}dk\Bigr) \nonumber \\
&-\text{I}\text{m}\Bigl(\int^{\pi}_{k_{\lambda}}\frac{2i (\cos
\alpha^{s}_{k})^{*}\sin \alpha^{s}_{k}\partial_{k}\alpha^{s}_{k}}{\cosh
2\text{I}\text{m}(\alpha^{s}_{k})}dk\Bigr) \nonumber \\
&-\text{I}\text{m}\Bigl(\int^{\pi}_{k_{\lambda}} \frac{2i|\cos \alpha^{s}_{k}|^{2} (\cos \alpha^{s}_{k} (\sin \alpha^{s}_{k})^{*}-\text{h}\text{.}\text{c}\text{.})\partial_{k} \alpha^{s}_{k}}{\cosh^{2} 2\text{I}\text{m}(\alpha^{s}_{k_{\lambda}})} dk \Bigr) \nonumber \\
&-\text{I}\text{m}\Bigl(\int^{k_{\lambda}}_{0} \frac{2i|\sin \alpha^{s}_{k}|^{2} (\cos \alpha^{s}_{k} (\sin \alpha^{s}_{k})^{*}-\text{h}\text{.}\text{c}\text{.})\partial_{k} \alpha^{s}_{k}}{\cosh^{2} 2\text{I}\text{m}(\alpha^{s}_{k_{\lambda}})} dk \Bigr).
\end{align}
Here the superscript $*$ denotes complex conjugation and h.c. denotes the hermitian conjugate.  We plot $C$ as a function of $s$ for different values of $\lambda$ in Fig.~\ref{fig:fig4}.  We recall that in Sec.~\ref{sec:GP}
using the ground state for the case of the TFIM~\cite{Heyl2012a} it was shown when
quenching within a phase led to a Chern number of $0$ while quenching across the critical point led to a change in topological quantization and $C = 1$.  That these values are integers is attributed to
the fact that the ground-state parameter manifold is topological
equivalent to a $S^2$-sphere, as discussed in Sec.~\ref{sec:GP}.  For the states we have studied, for $s \neq 0$ the
manifold of the states $\ket{s_{k,t}(\varphi)}$ does not possess the same properties as that of
$\ket{u_{k,t}(\varphi)}$.  These states depend non-trivially on $\varphi~
\forall~ (k \neq k_{\lambda})$, and so $M^2$ is cylindrical in nature.  Below $|\lambda| =
1$, the state defined by $k_{\lambda}$ does not depend on $\varphi$ and so the
cylinder is `pinched' at this point.  We illustrate these topologies in Fig.~\ref{fig:fig2}. 
These manifolds all possess boundaries and so display non-integer Chern numbers,
however on taking the limit of $s\rightarrow 0$ the manifolds become much
simpler.  Above $|\lambda| = 1$ they once again become equivalent to a
$S^2$-sphere and in the statically ordered phase, $|\lambda|<1$, all $\varphi$
dependence is lost.

The Chern number $C$ has a non-analytic point precisely at FCS critical line, see Fig.~\ref{fig:fig4}. This ``kink" is similar to that observed in the dynamical order parameter~\cite{Hickey2013} and examining the derivative $dC/ds$, we see it diverges at the critical line in a manner very similar to $\chi_{s}$.  We note that the FCS critical line is associated with a closing of the
gap of an \emph{excited state} in the complex spectrum of $H_{s}$~\cite{Hickey2013}. 
One may expect that the closing of such a gap would lead to divergences in the Berry phase and not extrema.  The emergence of extrema is due to our choice of initial state: while the gap in $H_s$ is single-particle in nature, the
multi-particle states $\ket{s}$ ($\ket{s_{t}}$) are of a construction such that in the
thermodynamic limit the single-particle divergences are suppressed and we see extrema.

\section{Conclusions}
\label{sec:Conc}

In this paper we examined singularities of the generating function of
time-integrated observables from the perspective of DPTs, as defined in~\cite{Heyl2013}, and geometric phase. 
We focussed on the example of the TFIM and the time-integrated transverse
magnetization.  Previously it was shown that using suitable external
environments one may prepare the system in specific states, $\ket{s}$, which
capture this FCS criticality.  We demonstrated, using such states,
that even \emph{far from quantum criticality} one may observe DPTs and that these
DPTs only emerge when one ``quenches'' across the FCS critical line. 
A recent work~\cite{Fagotti2013} highlighted a similar observation in the XXZ chain, where DPTs emerged when quenching within the gapped phase.  This was attributed to the steady state behaviour of the system.  In contrast in this paper we find that the emergence of DPTs is not due to the steady state behaviour but our \emph{initial state}. We then characterized this
FCS critical line by studying the geometry of these states.
We found that the derivative of the Berry phase with respect to the
$s$-parameter exhibited an extremum at the FCS critical line.  Similarly the
derivative of the Chern number diverged at this phase boundary.  These results
are similar to previous approaches to identifying static quantum critical points using
 the geometric phase.  This link between FCS criticality and the geometry of these $s$-states rquires
further investigation along with the examinstion of time-integrated
observables in other models to see if this connection still holds. Furthermore a direct link between
the generating functions of time-integrated observables and geometric quantities would be interesting and will be a focus of future work.

\section{acknowledgements}
This work was supported by EPSRC Grant no.~EP/I017828/1 and Leverhulme Trust
grant no.~F/00114/BG.

\appendix*

\section{Diagonalisation of $H_{s}/H$ and the $s$-state}
\label{app:diag}
To study the time-integrated magnetization of the TFIM we must diagonalize
the non-Hermitian Hamiltonian $H_{s}$, this can be done via a Jordan-Wigner
transformation in combination with a Bogoliubov rotation~\cite{Sachdev2011}. 
Then taking the $s \rightarrow 0$ limit one also obtains the diagonalized form
of the original Hamiltonian, $H$.

 The Jordan-Wigner transformation expresses the pauli spin operators
 $\sigma^{z}_{i}$, $\sigma^{+}_{i}$, and $\sigma^{-}_{i}$  at site $i$ in terms
 of corresponding fermionic operators ${c}_{i}$ and ${c}^{\dagger}_{i}$ with
 $\{{c}^{\dagger}_{i},{c}_{j} \}=\delta_{i,j}$ as
\begin{equation}
\label{eq:JWT}
\begin{split}
\sigma^{z}_{i} &= 1-2c^{\dagger}_{i}c_{i},\\
\sigma^{+}_{i} &= \prod_{j<i}(1-2c^{\dagger}_{j}c_{j}) c_{i}, \\
\sigma^{-}_{i} &= \prod_{j<i}(1-2c^{\dagger}_{j}c_{j}) c^{\dagger}_{i}.
\end{split}
\end{equation}
This Hamiltonian is translationally invariant and so we change to the Fourier
representation
\begin{equation}
c_{i} = \frac{1}{\sqrt{N}}\sum_{k}{e}^{-ikr_{i}}c_{k}
\end{equation}
and rewrite the Hamiltonian as
\begin{equation}
\label{eq:InterHam}
\begin{split}
H_{s} = \sum_{k}&\Bigl(2(\cos k-(is/2+\lambda))c^{\dagger}_{k}c_{k}\Bigr. \\
&\Bigl.-i\sin k
(c_{-k}c_{k}+c^{\dagger}_{-k}c^{\dagger}_{k}\Bigr).
\end{split}
\end{equation}
For specificity we restrict ourselves to an even number of spins $N$, and with
periodic boundary conditions, the discrete wave vector $k$ takes values $k =
\pi n/N$ where $n = -N+1, -N+3, \ldots,N-1$.

We note that the Hamiltonian in Eq.~(\eqref{eq:InterHam}) contains terms that do
not conserve the number of fermions, for instance
$c^{\dagger}_{-k}c^{\dagger}_{k}$. These terms are eliminated next via a canonical Bogoliubov rotation. This transformation expresses the Jordan-Wigner operators as a linear combination of a new set of $s$-dependent fermionic operators $c_{k}$ and $c^{\dagger}_{k}$ with $\{c_{k},c^{\dagger}_{k'}\}=\delta_{k,k'}$ as
\begin{equation}\label{eq:BogRot}
\begin{split}
c_{k} &= \cos \phi^{s}_{k}~A_{k} + i\sin \phi^{s}_{k}
~\bar{A}_{-k},\\
c^{\dag}_{k} &= \cos \phi^{s}_{k}~\bar{A}_{k} -
i\sin \phi^{s}_{k}~A_{-k}.
\end{split}
\end{equation}
Here we have a complex fermionic pair , $\{ \bar{A}_{k'},A_{k}\} =
\delta_{k',k}$, where $\bar{A}_{k} \neq A^{\dagger}_{k}$.  In the limit $s
\rightarrow 0$ this fermionic pair reduces to a more canonical form,
$\bar{A}_{k} \rightarrow \gamma^{\dagger}_{k}$ and $A_{k} \rightarrow
\gamma_{k}$. These complex Bogoliubov angles $\phi^{s}_{k}$ satisfy
\begin{equation}
\phi^{s}_{-k}=-\phi^{s}_{k}
\end{equation}
and are chosen such that only terms that conserve the number of fermions are
present in the Hamiltonian, note in the limit of $s\rightarrow 0$, this
angle becomes $\phi^{s=0}_{k} = \phi_{k}$. To enforce this condition, the Bogoliubov
angles must satisfy
\begin{equation}
\tan{\phi_{k}^{s}} =\frac{\sin{k}}{is/2+\lambda-\cos{k}}.
\end{equation}
With this choice we arrive at the free-fermion dispersion relation given:

\begin{equation}
\epsilon_{k}(\lambda,s) = 2 \sqrt{(\lambda + is/2 - \cos k)^2 + \sin^2 k}.
\end{equation}

Now the key property for expressing the fermionic states of the $s = 0$
Hamiltonian in terms of the fermionic states of $H_{s}$, is to notice that
ground state of $H$ may be express as a BCS state of $H_{s}$:

\begin{align}
\label{eq:BCS}
\ket{0}&= \frac{1}{\mathcal{N}'} \exp\left(\sum_{k>0}
B(k)\bar{A}_{k}\bar{A}_{-k}\right)\ket{0}_{s} \nonumber \\
&\propto \bigotimes_{k>0} \left[ \cos{{\alpha}_{k}^{s}}{\ket{0_{k},0_{-k}}}_{s}
-i\sin{{\alpha}_{k}^{s}}{\ket{1_k,1_{-k}}}_{s} \right]
\end{align}

In this Equation~\eqref{eq:BCS} $\ket{0_{k},0_{-k}}_{s}$ is the $k$-mode vacuum,
$A_{k}\ket{0_{k},0_{-k}}_{s}= A_{-k}\ket{0_{k},0_{-k}}_{s} = 0$, and 
$\bar{A}_{k}\bar{A}_{-k}\ket{0_{k},0_{-k}}_{s}=\ket{1_{k},1_{-k}}_{s}$ 
indicate occupation states of the fermionic modes with $|k|$ that diagonalise ${H}_{s}$.
  The complex angles appearing in the coefficients are directly related to the
 Bogoliubov angles by  $\alpha^{s}_{k} =
 \frac{\phi_{k}-\phi^{s}_{k}}{2}$.  
 This BCS form may be easily inverted and the fermionic occupation states,
 appropriately normalized, of $H_{s}$ are related to their $s = 0$ counterparts
 via:
\begin{align}
\ket{0_{k},0_{-k}}_{s}& = \frac{1}{\sqrt{\cosh 2\text{I}\text{m}
(\alpha^{s}_{k})}}\Bigl(\cos \alpha^{s}_{k}\ket{0_{k},0_{-k}} -i \sin
\alpha^{s}_{k}\ket{1_{k},1_{-k}}\Bigr), \nonumber \\
\ket{1_{k},1_{-k}}_{s} &= \frac{1}{\sqrt{\cosh 2\text{I}\text{m}
(\alpha^{s}_{k})}}\Bigl(\cos \alpha^{s}_{k}\ket{1_{k},1_{-k}} -i \sin
\alpha^{s}_{k}\ket{0_{k},0_{-k}}\Bigr).
\end{align}

We now consider the $s$-quench protocol described in the main text. We firstly
connect the Ising chain, prepared in its ground state, to a bath and allow it to
evolve to the $s$ state,

\begin{align}
\label{eq:sstate1}
\ket{s} &= \bigotimes_{k>0} \ket{s_{k}}, \nonumber \\
\ket{s_{k}} &= \tau(k_{\lambda}-k)\ket{0_{k},0_{-k}}_{s} +
\tau(k-k_{\lambda})\ket{1_{k},1_{-k}}_{s} \nonumber \\
&+\delta_{k,k_{\lambda}}\ket{0_{k_{\lambda}},0_{-k_{\lambda}}}.
\end{align}
here $\tau(x)$ is the
Heavi-side step function. 
After this we will evolve the state using it's original $s = 0$, Hamiltonian. 
Expressing these $H_{s}$ fermionic states in terms of their original $s = 0$
counterparts, this evolution may be evaluated analytically.
The time evolved $s$-state may still be expressed as a product of contributions from each momentum sector
\begin{align}
\label{eq:sstates}
&\ket{s_{k,t}} = \tau(k_{\lambda}-k)\frac{1}{\mathcal{N}}(\cos \alpha^{s}_{k} -
i \sin \alpha^{s}_{k}
{e}^{-2i\epsilon_{k}t}\gamma^{\dagger}_{k}\gamma^{\dagger}_{-k})\ket{0_{k},0_{-k}}
\nonumber \\
 &+ \tau(k-k_{\lambda})\frac{1}{\mathcal{N}}(-i \sin \alpha^{s}_{k} + \cos
\alpha^{s}_{k}
{e}^{-2i\epsilon_{k}t}\gamma^{\dagger}_{k}\gamma^{\dagger}_{-k})\ket{0_{k},0_{-k}}  \nonumber \\
 &+\delta_{k,k_{\lambda}} [\cos \phi_{k_{\lambda}} - i \sin \phi_{k_{\lambda}}
{e}^{-2i\epsilon_{k_{\lambda}}t}\gamma^{\dagger}_{k_{\lambda}}\gamma^{\dagger}_{-k_{\lambda}}]
\ket{0_{k_{\lambda}},0_{-k_{\lambda}}}
\end{align}
The normalization is related to the imaginary components of $\alpha^{s}_{k}$
given by $\mathcal{N} = \sqrt{\cosh 2\text{I}\text{m}(\alpha^{s}_{k})}$.  By
performing a global phase shift, $\gamma_{k}\rightarrow
\gamma_{k}{e}^{-i\varphi}$, we obtain $\ket{s_{k,t}(\varphi)}$ and
$\ket{s(\varphi)}$.

\bibliography{GEO}

\end{document}